\documentclass{article}

\usepackage{arxiv}

\usepackage[utf8]{inputenc} 
\usepackage[T1]{fontenc}    
\usepackage{hyperref}       
\usepackage{url}            
\usepackage{booktabs}       
\usepackage{amsfonts}       
\usepackage{nicefrac}       
\usepackage{microtype}      
\usepackage{lipsum}
\usepackage{graphicx}
\graphicspath{ {./images/} }

\usepackage{booktabs} 
\usepackage{tabularx}
\newcolumntype{C}{>{\centering\arraybackslash}X} 
\newcolumntype{L}{>{\arraybackslash}X} 
\usepackage{amsmath} 

\usepackage[ruled]{algorithm2e} 
\usepackage{array} 
\usepackage{longtable}   
\usepackage{pifont}  
\newcommand{\cmark}{\ding{51}}%

\title{Dynamic detection of mobile malware using smartphone data and machine learning}

\author{
 J.S. Panman de Wit \\
  University of Twente\\
 \And
 J. van der Ham \\
  University of Twente\\
 \And
 D. Bucur \\
  University of Twente\\
}

\begin{document}
\maketitle
\begin{abstract}
Mobile malware are malicious programs that target mobile devices. They are an increasing problem, as seen in the rise of detected mobile malware samples per year. The number of active smartphone users is expected to grow, stressing the importance of research on the detection of mobile malware. Detection methods for mobile malware exist but are still limited. 

In this paper, we provide an overview of the performance of machine learning (ML) techniques to detect malware on Android, without using privileged access. The ML-classifiers use device information such as the CPU usage, battery usage, and memory usage for the detection of 10 subtypes of Mobile Trojans on the Android Operating System (OS).

We use a real-life dataset containing device and malware data from 47 users for a year (2016). We examine which features, i.e. aspects, of a device, are most important to monitor to detect (subtypes of) Mobile Trojans. The focus of this paper is on dynamic hardware features. Using these dynamic features we apply state-of-the-art machine learning classifiers: Random Forest, K-Nearest Neighbour, and AdaBoost. We show classification results  on different feature sets, making a distinction between global device features, and specific app features. None of the measured feature sets require privileged access.

Our results show that the Random Forest classifier performs best as a general malware classifier: across 10 subtypes of Mobile Trojans, it achieves an F1 score of 0.73 with a False Positive Rate (FPR) of 0.009 and a False Negative Rate (FNR) of 0.380. The Random Forest, K-Nearest Neighbours, and AdaBoost classifiers achieve F1 scores above 0.72, an FPR below 0.02 and, an FNR below 0.33, when trained separately to detect each subtype of Mobile Trojans. 
\end{abstract}

\section{Introduction}

Nowadays smartphones have become an integral part of life, with people using their phones in both their private and professional life. The number of active smartphone users globally is expected to be 7.3 billion  by 2025 \cite{statista2021}. The rise in smartphone users has also led to an increase in malicious programs targeting mobile devices, i.e. mobile malware. Criminals try to exploit vulnerabilities on smartphones of other people for their own purposes. Additionally, over the past years, malware authors have become less recreational-driven and more profit-driven as they are actively searching for sensitive, personal, and enterprise information \cite{tam2017evolution}.

Detection of mobile malware is becoming increasingly relevant, with machine learning showing the most promise. In this paper we present an overview of the performance of state-of-the-art machine learning classifiers using sensor data on the Android mobile operating system. We show different training approaches for each of the classifiers, showing their performance on individual and general cases. These machine learning classifiers can be trained on a cluster on a large dataset, resulting in a classifier. This classifier can then be deployed on mobile phones to detect future malware based on the dynamic sensor data similar to the dataset, i.e. dynamic analysis.

Academic work is mainly divided into dynamic analysis and static analysis of mobile malware. Dynamic analysis refers to the analysis of malware during run-time, i.e. while the application is running. Static analysis refers to the analysis of malware outside run-time, e.g. by analysing the installation package of a malware app. Dynamic analysis has advantages over static analysis but methods are still imperfect, ineffective, and incomprehensive \cite{yan2017survey}.

An important limitation is that in most studies of malware detection, virtual environments are used, e.g. analysis on a PC, instead of real mobile devices. An increasing trend is seen in malware that use techniques to avoid detection in virtual environments, thereby making methods based on analysis in virtual environments less effective than methods based on analysis on real devices \cite{tam2017evolution}. Moreover, we found that most methods are assessed with i) malware running isolated in an emulator, and ii) malware running for a brief period. This kind of assessment does not reflect the circumstances of a real device with for example different applications running at the same time. Therefore, most research does not provide a realistic assessment of detection performances of their detection methods due to their unrealistic circumstances.

Using a dataset \cite{sherlock} containing data of 47 users throughout the year 2016, this research provides valuable knowledge on detecting mobile malware on real devices and a realistic assessment of the research's detection methods. This paper focuses on dynamic hardware features as our initial literature exploration showed relatively few research compared to other types of dynamic features (e.g. software and firmware). In this work, we compare the detection performance of the Random Forest, K-Nearest Neighbour, and AdaBoost classifiers. The classifier performance is assessed on 10 subtypes of Mobile Trojans, i.e. malware showing benign behaviour but performing malicious behaviour, as these are the most prevalent mobile malware type \cite{googleReport2017,kasperskyReport2017}. Furthermore, we examine which features are important in detecting different subtypes of Mobile Trojans. We find that the Random Forest and AdaBoost classifiers have similar and better performances than the K-Nearest Neighbour. Additionally, our initial results show that AdaBoost may be more suitable than the other classifiers, for use cases where detection models are built with data from one subset of users and subsequently used for protecting other users. Additionally, we show that application-specific features are important in detecting mobile malware. In contrast to other literature, which often assessed detection models on devices running one or few applications in virtual environments, we find low predictability using only global device features. 

In summary, we make the following contributions:
\begin{itemize}
\item We use a recent real-life sensor dataset to create multiple mobile malware detection models. Most literature is based on environments with models based on emulator data, simulated user input, and/or short testing durations. Our research takes a realistic approach, contributing to the body of scientific literature on malware detection that is based on real-life data rather than data derived from emulators and virtual environments.  
\item Our results are based on a dataset containing data from 47 real users over a time span of 1 year. To the authors' best knowledge, this is one of the first research on data from this many users over a long time period. The large user pool and time span increases the confidence in the generalisability of the results.
\item We evaluated our detection methods on a Random Forest, K-nearest neighbour and AdaBoost classifiers with several performance metrics. AdaBoost has been researched little in literature. We provide the performance of the classifiers with several training and testing methods. 
\item Performance of the classifiers is shown for seperate mobile malware types. The results show a large difference in performance for several mobile malware types, indicating the difficulty of detecting some mobile malware types. Earlier research, to the best of our knowledge, did not include a comparison of this many mobile malware types.
\item We briefly describe a testing method where models are trained on data of one set of users and tested on data from another set of users. This testing method may be useful when detection models are built with one user pool and are subsequently used to protect another, perhaps larger, user pool.
\item All features included in this research do not require any privileged access. This increases the usability of mobile applications based on our detection methods. Additionally, we show which feature categories and which specific features are most important in detecting different malware types.
\end{itemize}

\section{Related works}\label{sec:related_works}

In an early, highly cited study (Shabtai ~\cite{shabtai2012andromaly} from 2011), the authors designed a behavioural malware detection framework for Android devices called Andromaly. This used device data: touch screen, keyboard, scheduler, CPU load, messaging, power, memory, applications, calls, processes, network, hardware, binder, and LED. 40 benign applications and 4 self-developed malware applications (a DOS Trojan, SMS Trojan, Spyware Trojan, and Spyware malware) were executed. When training the detection model on data from the same mobile device, the J48 decision tree classifier performed best; on the other hand, a Naive Bayes classifier detected malware best in experiments in which the testing was done on data from a new device (see the overview in Table~\ref{tab:related_works}). Despite relatively high False Positive Rates with all their models, Andromaly showed the potential of detecting malware based on dynamic features using machine learning, compared different classifiers, trained and tested on data collected from real devices. However, the study is relatively old, and the malware ecosystem has since changed.

\begin{table}[htb]
\caption{Related work. When multiple malware classifiers are studied, the best one is emphasized.}
\caption*{\linespread{0.5}\selectfont{} {\scriptsize 
Legend: 
\textit{Acc} accuracy, 
\textit{AZ} Androzoo \cite{allix2016androzoo},
\textit{bat} battery, 
\textit{BN} Bayesian Network, 
\textit{Cr} Crowdroid \cite{crowdroid}, 
\textit{cust} Custom, 
\textit{Dr} Drebin \cite{drebin}, 
\textit{DS} Decision Stump, 
\textit{DT} Decision Tree, 
\textit{GBT} Gradient Boosted Tree
\textit{Ge} Malware Genome Project \cite{malware_genome}, 
\textit{GM} Gaussian Mixture,
\textit{histo} Histogram, 
\textit{IR} Isotonic Regression
\textit{Kmeans} K-means clustering, 
\textit{KNN} K-Nearest Neighbour, 
\textit{LDA} Linear Discriminant Analysis,
\textit{LDC} Local Density Cluster-Based Outlier Factor, 
\textit{loc} Location, 
\textit{LR} Logistic Regression, 
\textit{mem} memory, 
\textit{md} metadata, 
\textit{MLP} Multi-Layer Perceptron, 
\textit{mult} Multiple datasets,
\textit{NB} Naive Bayes, 
\textit{netw} network, 
\textit{NN} Neural Network, 
\textit{Om} Open Malware \cite{openmalware},
\textit{perm} permissions, 
\textit{PC} Parzen classifier
\textit{proc} process
\textit{PS} Play Store, 
\textit{QDA} Quadratic Discriminant Analysis,
\textit{RBF} Radial Basis Function
\textit{RF} Random Forest, 
\textit{sens. act} Sensitive activities,
\textit{ShD} Sherlock dataset
\textit{stdev} standard deviation, 
\textit{sto} storage, 
\textit{SVM} Support Vector Machine, 
\textit{SC} system calls, 
\textit{UP} user presence, 
\textit{VE} Virtual Environment, 
\textit{VS} VirusShare~\cite{virusshare}, 
\textit{VT} VirusTotal~\cite{virustotaldb}}}
\label{tab:related_works}
\begin{minipage}{\columnwidth}
\begin{center}
\resizebox{1.\textwidth}{!}{\begin{tabular}{m{.4cm} m{.6cm} >{\raggedright\let\newline\\\arraybackslash\hspace{0pt}}m{2.8cm} m{.75cm} m{.95cm} m{1.25cm} m{1.8cm} m{2.5cm} m{.55cm} m{.55cm }m{.55cm}}
\toprule
\multicolumn{2}{c}{\textbf{Article}} & \multicolumn{2}{c}{\textbf{Features}} & \multicolumn{3}{c}{\textbf{Training and testing}} & \multicolumn{4}{c}{\textbf{Performance}}\tabularnewline
\cmidrule(lr){0-1}\cmidrule(lr){3-4}\cmidrule(lr){5-7}\cmidrule(lr){8-11}
\textbf{Ref}  & \textbf{Year}  & \textbf{Dynamic}  & \textbf{Static}  & \textbf{Benign}  & \textbf{Malware}  & \textbf{Platform}  & \textbf{Classifiers}  & \textbf{Acc}  & \textbf{TPR}  & \textbf{FPR}\tabularnewline
\cmidrule(lr){0-1}\cmidrule(lr){3-4}\cmidrule(lr){5-7}\cmidrule(lr){8-11}
\textbf{\cite{shabtai2012andromaly}}  & \textbf{2012}  & various (14)  & - & 40  & 4$^{\text{cust}}$  & 2 devices  & BN, histo, \textbf{J48}, Kmeans, LR, \textbf{NB}  & 0.81  & 0.79  & 0.48 \tabularnewline
\textbf{\cite{amos_applying_2013}}  & \textbf{2013}  & bat, Binder, CPU,\newline mem, netw  & perm  & 408$^{\text{PS}}$  & 1330$^{\text{Ge,VT}}$  & VE, Monkey  & \textbf{BN}, J48, LR, MLP, NB, RF  & 0.81  & 0.97  & 0.31 \tabularnewline
\textbf{\cite{alam_random_2013}}  & \textbf{2013}  & Binder, CPU, mem  & - & 408$^{\text{PS}}$  & 1130$^{\text{Ge,VT}}$  & VE, Monkey  & \textbf{RF}, BN, NB, MLP, J48, DS, LR  & 1.00  & -  & few \tabularnewline
\textbf{\cite{ham2013analysis}}  & \textbf{2013}  & CPU, net, mem, SMS  & - & 30$^{\text{PS}}$  & 5$^{\text{cust}}$  & 1 device  & NB, \textbf{RF}, LR, SVM  & -  & 0.99  & 0.00 \tabularnewline
\textbf{\cite{attar_gaussian_2014}}  & \textbf{2014}  & bat, CPU, mem, netw  & - & >{$2^{\text{PS}}$}  & 3$^{\text{Cr}}$  & 12 devices  & GM-LDC  & $\approx$1  & $\approx$1  & $\approx$0\tabularnewline
\textbf{\cite{dixon_power_2014}}  & \textbf{2014}  & bat, time, loc  & - & - & 2$^{\text{cust}}$  & 11 devices  & stdev  & $\approx$1  & -  & $\approx$0 \tabularnewline
\textbf{\cite{saracino2016madam}}  & \textbf{2016}  & SC, SMS, UP  & md  & {$9804^{\text{PS}}$}  & 2800  & 3 devices  & 1-KNN, LDA, QDC, MLP, PC, RBF & -  & 0.97  & 0.00 \tabularnewline
\textbf{\cite{caviglione_seeing_2016}}  & \textbf{2016}  & bat  & - & - & 7$^{\text{cust}}$  & 1 device  & \textbf{NN}, DT  & -  & \textgreater0.85  & -\tabularnewline
\textbf{\cite{milosevic2016malaware}}  & \textbf{2016}  & CPU, mem  & - & 940$^{\text{PS}}$  & 1120$^{\text{Ge}}$  & VE, Monkey  & LR  & -  & 0.86  & 0.17 \tabularnewline
\textbf{\cite{ferrante2016spotting}}  & \textbf{2016}  & CPU, mem, SC  & - & 1709$^{\text{PS}}$  & 1523$^{\text{Dr}}$  & VE, Monkey & Kmeans-RF & 0.67  & 0.61 & 0.28 \tabularnewline
\textbf{\cite{canfora2016acquiring}}  & \textbf{2016}  & CPU, mem, net, sto & - & 1059$^{\text{PS}}$  & 1047$^{\text{Dr}}$  & VE, Monkey  & RF  & 1.00  & 0.82  & 0.01 \tabularnewline
\textbf{\cite{massarelli2017android}}  & \textbf{2017}  & CPU, mem, net  & - & - & \textless 5560$^{\text{Dr}}$  & VE, Monkey & SVM & 0.82 & - & - \tabularnewline
\textbf{\cite{wassermann2018bigmomal}}  & \textbf{2018}  & CPU, net  & - & \textgreater 10 & 3$^{\text{ShD}}$  & 47 devices & DT & 1 & 1 & \textless 1\% \tabularnewline
\textbf{\cite{cai2018droidcat}} & \textbf{2018} & Method calls, Intents & - & 17k$^{\text{PS, AZ}}$ & 13.9k$^{\text{mult}}$ & VE, Monkey & RF    & 0.98 & 0.98  & - \tabularnewline
\textbf{\cite{martin2018candyman}} & \textbf{2018} & Network, SMS, Files, more &     -  &     -  & 4442$^{\text{Dr}}$ & VE, Monkey & \textbf{RF}, DT, KNN, SVM, NN & 0.813 & 0.813 & - \tabularnewline
\textbf{\cite{dai2019smash}} & \textbf{2019} & CPU, mem, api & -     & -     & 27000$^{\text{Om}}$ & VE    & SVM, RF, \textbf{NN} & 0.97  & -     & - \tabularnewline
\textbf{\cite{memon2019comparison}}  & \textbf{2019}  & Bat, CPU, net, proc  & - & 3$^{\text{ShD}}$ & 3$^{\text{ShD}}$  & 47 devices & IR, RF, DT, \textbf{GBT}, MLP, SVM, LR & 1 & 1 & \textless 1\% \tabularnewline
\textbf{\cite{alzaylaee2020}} & \textbf{2020} & API calls & perm  & 11505 & 19620 & 8 devices, DynaLog & \textbf{NN}, NB, SVM, DT & 0.95 & 0.98  & 0.09 \tabularnewline
\midrule 
this &  & CPU, bat, mem, net, sto & - & 10$^{\text{cust}}$ & 10$^{\text{cust}}$ & 47 devices  & \textbf{RF}, NB, KNN, MLP, AdaBoost & 0.96  & 0.65  & 0.01 \tabularnewline
\bottomrule
\end{tabular}}
\end{center}
\end{minipage}
\end{table}

The STREAM framework \cite{amos_applying_2013}, aims to enable rapid large-scale validation of machine-learning classifiers for mobile malware. STREAM used 41 features (listed in Table~\ref{tab:related_works}) collected every 5 seconds from emulators running in a so-called ATAACK cloud. The emulator used the Android Monkey application to simulate pseudo-random user behaviour such as input on the touchscreen. To evaluate their detection model, the authors used six classifiers, with good accuracy and TPR for the Bayesian Network. The training set had 408 popular applications from the Google Play Store and 1330 malware applications from the Malware Genome Project database~\cite{malware_genome} and the VirusTotal database~\cite{virustotaldb}. The testing set had 24 benign and 23 malware applications from the same databases. This showed the potential of using dynamic features, although the FPR for all classifiers was relatively high. Additionally, this research ran applications separately for 90 seconds and made use of a virtual environment in the form of an emulator with user-like behaviour created by the Android Monkey tool. This lowers the confidence that the model would perform the same in real life. The same dataset from Amos \cite{amos_applying_2013} was used in Alam \cite{alam_random_2013} for anomaly detection with application behaviour features: only the Binder, CPU, and memory features. The dataset was then balanced. 
The Random Forest classifier had an excellent accuracy of 99.9857\% and a root MSE of 0.0183\%. Only 2 false positives were measured during this experiment. However, it is sensitive to the same limitations as Amos \cite{amos_applying_2013}.

Saracino \cite{saracino2016madam} describes an improvement of an earlier anomaly detection model (Dini \cite{dini2012madam}). This version includes system calls, user presence, and SMS features, plus a static analysis of application packages. Many classifiers were trained and tested. 
Their performance is not reported, but the authors state that 1-Nearest Neighbour achieved the best classification results. To evaluate the TPR, the detection framework has been tested against 2800 malware applications from the Malware Genome Project, Contagio~\cite{contagio}, and VirusShare~\cite{virusshare}, representing the Botnet, Installer, Trojan, Ransomware, Rootkit, SMS Trojan, and Spyware malware families. How long the different malware applications were run is not mentioned. Also, it is unknown whether the same device was used for the training and testing of the classifiers. The framework achieved very good TPR and FPR, and also detected some zero-day attacks, undetected by multiple Antivirus Software at that time. 
The performance was assessed on real devices, varying the usage intensity over a period of one week, to reflect real-life circumstances. A limitation is a requirement for root permissions, as system calls were used as features.

In Milosevic \cite{milosevic2016malaware}, memory and CPU usage are predictive features, tracked by running every application separately for 10 minutes in an Android emulator. The emulator was fed with user-like input by the Monkey application. Their optimized feature set only contained 7 features. The classification algorithm used was linear Logistic Regression with the use of a sliding window technique. This was validated on a set of 94 benign and 89 malware applications.

The classifier with the highest TPR of 95.7\% of malware had a relatively high FPR of 25\%; the classifier with the highest F-measure had a TPR 85.5\% and an FPR of 17.2\%. This study showed the potential of using the memory and CPU usage as features for the dynamic detection of mobile malware. However, the detection had relatively high FPR, besides the disadvantage of using emulator data.

More studies were based on emulation. In Ferrante \cite{ferrante2016spotting}, the features were related to system calls, memory, and CPU usage. Benign apps were downloaded from the Google Play store and malicious applications were collected from the Drebin dataset~\cite{drebin}. The applications were run for 10 minutes in an Android emulator and features were collected every 2 seconds. The emulator was fed with user-input from the Monkey application. As detection models, they first used a K-means clustering algorithm to cluster apps based on the similarity of memory and CPU usage, then a Random Forest classifier on every cluster that classified the applications based on their system calls. The best performing model was the classifier with 7-means clustering and a Random Forest classifier of 50 trees. This showed the potential of using system calls, memory, and CPU features for malware detection. The performance is relatively low compared to other detection methods, and the method needs root permission as it uses system calls as features. In Canfora \cite{canfora2016acquiring}, the features are related to CPU, memory, storage and network. The applications were run for 60 seconds in an Android emulator. The features were used both raw and after a Discrete Cosine Transformation. Furthermore, the authors varied the granularity of the detection method by either taking all the features, only the global features, or only the features of the application under analysis. The Random-Forest classifier using global features had very good accuracy and FPR. 

In Massarelli \cite{massarelli2017android} the authors used the Drebin dataset for malicious applications (but it is unknown how many malware applications were included from the 5560 in the dataset). The features were both system-wide and application-specific. An emulator was used for  training the model, fed with simulated user input events from the Monkey application. The classifier was an SVM with a Radial Basis Function kernel, which had good accuracy. The exact FPR is unknown, but the precision of the model ranges from 10\% to 90\% depending on the malware family, which is low in comparison with other methods. This also used an emulator, leading to limitations in its applicability.

In Martin \cite{martin2018candyman}, multiple classifiers, including Random Forest, K-nearest neighbour, and SVMs are assessed. They achieve an F1 score of 0.803 on a dataset containing 4442 malicious malware samples. They use hardware features related to network, SMS, file reads/writes. Other features included are:  services started, classes loaded, information leaks via the network, file and SMS services, circumvented permissions, cryptographic operations, SMSs sent, and phone calls. Their detection models are also assessed using a virtual environment rather than devices. This is also the case for Dai \cite{dai2019smash}. They use features related to CPU, Memory and API calls (but also software features such as API calls, which this paper does not). Their best classifier is well-performing with an F1-score of 0.98. Additionally, they assess their detection model on a large dataset of more than 27.000 malicious applications.

During the writing of this paper, other authors used the SherLock dataset, but to a much more limited extent than this research.
In Memon \cite{memon2019comparison}, 7 classifiers were trained for malware detection: Isotonic Regression, Random Forest, Decision Trees, Gradient-Boosted Trees, MultiLayer Perceptron, SVM, and Logistic Regression. The data was balanced so that 50\% of the labels were benign and 50\% were malicious. Only data from Q3 2016 was studied, meaning that only three malware types (Madware, Ransomware and Click-jacking) were included for training and testing, with CPU, network traffic, battery, and process features. The Gradient-Boosted Trees had the best results with an F1 score of 0.91\% and an FPR of 0.09\%. Their Random Forest classifier also worked almost equally well. These make for better results than prior work, which may be due to the limited number of malware types included. 

In Wassermann \cite{wassermann2018bigmomal}, Decision Trees were trained as classifiers, only on SherLock data from Q2 2016, so only three malware types (Spyware, Phishing, Adware). The features used were network traffic and CPU-related. The label to predict was whether the malicious app (Moriarty) was running. Their model results in a recall of almost 100\% with less than 1\% FPR. This research showed good app-detection results, although on a limited set of malware; also, the model cannot distinguish between individual (benign or malicious) actions; it only predicts the presence of the malware on the system.

Recent literature on dynamic malware detection (Droidcat  \cite{cai2018droidcat} and Dl-Droid \cite{alzaylaee2020}) have high performance on a large dataset containing more than 10000 malware applications. However, DroidCat focuses on software features such as Method Calls and trained and tested their method within a virtual environment with simulated pseudo-random user behaviour. Dl-Droid achieves a TPR of 0.95 and an FPR of 0.09, but also only including software features such as API calls. This paper focuses on hardware features only. 

Recent work by (Cai et. al. \cite{cai2020longitudinal}) analysed 17.664 Android applications developed throughout 2010--2017. The paper describes differences in method calls, ICC calls and source/sink calls during static-code analysis and dynamic analysis. One of the finding of the paper is that access to sensitive data was almost 10x more frequent during dynamic analysis than during static analysis (Figure 17 and Finding 8). This finding indicates that dynamic analysis may support in capturing all sensitive data calls and could therefore help in detecting mobile malware.

Other research assessed have also used machine learning for the dynamic detection of mobile malware, but this literature is unclear about the feature collection method (Ham and Choi \cite{ham2013analysis}), include few malware samples (Attar \cite{attar_gaussian_2014}, Dixon \cite{dixon_power_2014}), or experiment on an idle phone (Caviglione \cite{caviglione_seeing_2016}).

While we have seen a body of prior work using dynamic hardware features to detect malware on mobile devices, important issues remain unaddressed. Most recent papers have trained their detection model on emulators with simulated input; this makes it difficult to extrapolate their performances with real users and to have high confidence in how realistic the results are. Additionally, most studies have only executed the benign and malicious applications for a few minutes, meaning that any delayed malicious behaviour would not be detected. Also, a limited amount of research has shown excellent results, particularly in what regards the FPR metric. In this paper, we address these issues and provide a more realistic picture by learning from data gathered with real users, from a study which executed malware from different categories, over a large time span. 

\section{Method}\label{sec:method}

\subsection{Malware data}

We use the SherLock dataset \cite{sherlock} provided by Ben-Gurion University. This dataset contains 10 billion data records in system logs collected from 47 Samsung Galaxy S5 devices used by 47 different volunteer users, over a period of over one year (2016). During this year, the volunteers executed both benign and malicious applications on their devices. The data collection agent on the phone is based on the Funf Open Sensing framework, developed by MIT Media Lab \cite{google_funf}. This framework allows for the collection of sensor data (e.g. memory consumption or CPU usage). The authors adjusted the source code to improve stability and reliability. Additionally they added features such as the collection of statistics on all running applications of a smartphone. According to the SherLock paper\cite{sherlock} all features can be tracked without any root permissions.  The source code of the data collection agent can be found on GitHub \cite{github_sherlock}.

The malicious applications were written by the researchers based on wild malware. Every month a different subtype of Mobile Trojan was installed on the devices. Each malware version resembled a subtype of Mobile Trojan with both benign and malicious actions. In total, 11 malware versions were included, listed in Table \ref{tab:malware_versions} with their benign behaviour, malicious behaviour (as described by \cite{sherlock}), malware type, a description of their actions, and the wild malware on which the implementation was based. Important to note is that each time data was transmitted by the malware application, it was scrambled prior to sending to protect the privacy of the users.

\begin{table}[htb]
    \caption{Malware types}
    \label{tab:malware_versions}
    \begin{minipage}{\columnwidth}
    \begin{center}
        \resizebox{1.\textwidth}{!}{
\begin{tabular}{m{0.5cm} >{\raggedright\let\newline\\\arraybackslash\hspace{0pt}}m{2cm} >{\raggedright\let\newline\\\arraybackslash\hspace{0pt}}m{2.2cm} m{1.8cm}m{7.2cm} >{\raggedright\let\newline\\\arraybackslash\hspace{0pt}}m{2cm}}
\toprule
 & {\bf Benign behaviour} & {\bf Malicious\newline behaviour} & {\bf Type} & {\bf Description} & {\bf Wild\newline malware} \\
\hline
1 & Game & Contact theft & Spyware & Steals, encrypts, and transmits all contact stored on device. & SaveMe, SocialPath\\
2 & Web browser & Spyware & Spyware & i) Spies on location and audio, or ii) spies on web traffic and web
history. & Code4hk, xRAT\\
3 & Utiliz. Widget & Photo theft & Spyware & Steals photos that are taken and in storage, and takes candid photos
of the user. & Photsy, Phopsy\\
4 & Sports app & SMS bank thief & SMS Fraud & Captures and reports immediately on SMSs that contain codes and various
keywords. & Spy.Agent.SI\\
5 & Game & Phishing & Phishing & Makes fake shortcuts and notifications to login to Facebook, Gmail, and Skype. & Xbot\\
6 & Game & Adware & Adware & Gathers information and places ads, popups and banners. & -\\
7 & Game & Madware & Hostile downloader & Gathers private information, places shortcuts, notifications, and
attempts to install new applications. & -\\
8 & Lock-screen & Ransomware & Ransomware & Performs either: 1) lock screen ransomware, or 2) crypto ransomware. & Simplocker.A, SLocker\\
9 & File Manager & Click-jacking & Privilege\newline escalation & Tricks the user to activate accessibility services to then hijack the user interface. & Shedun (GhostPush)\\
10 & - & Device theft & - & - & \\
11 & Music Player & Botnet & DOS & Either performs: 1) DDoS attacks on command, or 2) SMS botnet activities & Tascudap.A, Nitmo.A\dots{}\\
12 & Web media\newline player & Recon. Infiltr. & Other & Maps the connected local network and searches for files and vulnerabilities. & -\\
\bottomrule
\end{tabular}

}
    \end{center}
    \end{minipage}
\end{table}

The dataset is divided into 13 probes, namely groups of multiple sensors that shared the same sample interval. This research uses the following probes:
\begin{itemize}
    \item the Moriarty (\emph{Malware}) probe for malware data,
    \item the T4 (\emph{System}) probe for global device data, and
    \item the Applications (\emph{Apps}) probe for app-specific device data.
\end{itemize}

The Malware probe sensed data from the malicious application installed on the devices. Each malware probe's record is a log of the \emph{action} taken by the user's malware application. Each action is part of a malware \emph{session}: the malware application started in either a benign or malicious session. Within a benign session, only benign actions were performed; in a malicious session, the malware performed both benign and malicious actions. An overview of Malware probe's columns, their data type, and a description is shown in Table \ref{tab:moriarty_data_overview}. For this research, we excluded version 10 and 12 as no malicious data was available for these versions.

\begin{table}[htb]
    \caption{Malware probe data overview}
    \label{tab:moriarty_data_overview}
    \begin{minipage}{\columnwidth}
    \begin{center}
        {\small
        \begin{tabular}{lll}
\toprule
\textbf{Column} & \textbf{Datatype} & \textbf{Description}
\tabularnewline
\midrule
UserId & String & User ID \\
UUID  & Numeric & Timestamp of action \\
Details & String & Details of action \\
Action & String & Action performed \\
ActionType & String & Whether action is benign or malicious \\
SessionType & String & Whether session is benign or malicious \\
Version & Decimal & Version number \\
SessionID & Decimal & Session ID \\
Behavior & String & Behaviour of current session \\

\bottomrule
\end{tabular}
        }
    \end{center}
    \end{minipage}
\end{table}

The System probe tracked global device data every 5 seconds. Each record of the system probe is a log of the user's global device data at a given time, taken from the \texttt{/proc/} folder. The feature categories that were tracked are battery, CPU, network, memory, I/O interrupts, and storage. 41 features are used from this probe.

The App probe recorded app data every 5 seconds for each application installed on the device. For this research, the only relevant data is the app data of the malware application. Each record is a log of the malware app data at a given time, taken from the \texttt{/proc/} folder. 35 features are used from this probe. For a complete overview of the features used, refer to Table  \ref{tab:features_used_overview}.

The data from the three probes is integrated by joining on the \textit{userid} and \textit{timestamp} fields. Not all probes had the same tracking frequency so joining on the timestamp column can not be done with a precise join. Therefore, probe data with a maximum of 5 seconds after the timestamp of the Malware data is used in the joining of the data. This way we can match 83\% of the Malware data with probe data. Additionally, we apply random undersampling to rebalance the original dataset. The original dataset contains 90\% malicious records and only 10\% benign records. The dataset is balanced by downsizing the number of malicious data points per malware type until 90\% of the rows are benign actions and 10\% are malicious. The malicious data points that are removed are chosen uniformly at random. The undersampling method is applied to both the training and testing data. We apply the resampling method on the training data to prevent the classifiers from a bias towards the (original) majority class (i.e. malware data). We apply the resampling method on the testing data to to better reflect real-life circumstances and to provide more realistic classification results. In the real world, the number of malicious actions on a mobile phone would be relatively low compared to the number of benign records. Lastly, we normalize any features in the dataset by scaling it between zero and one for the KNN classifier. No data normalization is used for AdaBoost and RF as these are tree-based classifiers. The dataset now consists of 28821 rows, 102 columns, and is 21.4 MB.

\subsection{Supervised learning algorithms}

We train a statistical classifier able to recognize malware signatures in any log data collected on a smartphone. The classifier is trained, cross-validated, and tested using the dataset described above. Three training algorithms yielded well-performing classifiers: \emph{Random Forest} and \emph{K-Nearest Neighbour} classifiers had a good performance in our related work (see  Section \ref{sec:related_works}), so we include them in our study. We add the \emph{AdaBoost} classifier, which is designed specifically to improve the performance of a Random Forest. All are nonlinear classifiers, able to capture complex relationships between the variables in the dataset. Table~\ref{tab:classifier_pros_cons} (based on \cite{cybersecurityDataMining}) lists advantages and disadvantages for these algorithms. 

\begin{table}[htb]
    \caption{Pros and cons of classifiers}
    \label{tab:classifier_pros_cons}
    \begin{minipage}{\columnwidth}
    \begin{center}
        {\small

\begin{tabularx}{.9\textwidth}{lll}
\toprule
\textbf{Classifier} & \textbf{Pros} & \textbf{Cons}\tabularnewline
\hline
\textbf{Random Forest (RF)} & Can overcome overfitting & A greedy algorithm \tabularnewline
\hline
\textbf{AdaBoost} & Nonlinear classification & Sensitive to imbalanced dataset \tabularnewline
 & Efficient with high-dimensional data &  \tabularnewline
\hline
\textbf{K-Nearest} & High precision and accuracy & Computationally expensive\tabularnewline
\textbf{Neighbours (KNN)} & Nonlinear classification & Sensitive to imbalanced dataset\tabularnewline
 & No assumption of features & \tabularnewline
\bottomrule
\end{tabularx}

        }
    \end{center}
    \end{minipage}
\end{table}

\subsection{Training, cross-validating, and testing malware classifiers}

All classifiers are trained to identify a malicious action (the ActionType column of the Malware dataset). The classifiers that are trained differ in their input data (feature set), classification target, and test mode. Three feature sets are assessed: Global, Apps and a combination of both. The global feature set includes global device features from the System probe. The apps feature set include features from the Apps probe. 

Two classification targets are tried. The first classification target (all malware) trains one general malware classifier to identify malicious actions by any of the 10 malware types in the dataset. The second classification target (single malware) trains one single-purpose classifier per malware type, resulting in 10 different trained models.

Two test modes are run. Both use a percentage of the data for training, and the remaining for testing. In the \emph{normal holdout test mode}, the training and testing data is sampled at random. In the \emph{unknown device test mode}, the testing data contain data of devices that are not in the training data. The unknown device test mode is used to simulate a situation in which no data is yet available for a certain device. In that case, a model trained on data from other devices is a solution. Testing the models on other devices also helps to detect the potential bias of the model on a specific device. Note that only the testing on the test set is adjusted; the cross-validation remains the same.

\begin{table}[htb]
    \caption{Different configurations for the classifiers trained}
    \label{tab:experiment_variables}
    \begin{minipage}{\columnwidth}
    \begin{center}
        {\small 

\begin{tabularx}{0.85\textwidth}{lll}
\toprule
 & \textbf{Type} & \textbf{Description}  \tabularnewline
\hline
\textbf{Feature set} & Global & Device features (the System probe) \tabularnewline
 (complete listing in Table~\ref{tab:features_used_overview}) & Apps & App-specific features (the Apps probe) \tabularnewline
 & Combined & Both Global and Apps features \tabularnewline
\hline
\textbf{Classification target} & All malware & Generic classifier across malware types \tabularnewline
 & Single malware & One classifier for each malware type \tabularnewline
\hline
\textbf{Test mode} & Normal holdout & Training data includes all devices \tabularnewline
 & Unknown device & Test set includes data from new devices \tabularnewline
\bottomrule
\end{tabularx}

        }
    \end{center}
    \end{minipage}
\end{table}

To train the dataset, we use 4-fold cross-validation with holdout. The hyperparameters of each classifier are tuned using a grid search. The search space for the hyperparameter settings are as follows. For AdaBoost, the number of estimators is in the interval [5, 400]; for Random Forest, the number of trees in the interval [5, 320], the maximum tree depth in [3, 320], the maximum number of features in [3, 102]; for K-Nearest Neighbours, the number of neighbours in [1, 61]. We also use Recursive Feature Elimination together with cross-validation (RFCV) to find the optimal number of features for all classifiers. RFCV requires a feature ranking, as the method recursively eliminates the least important feature. The feature rankings are based on the Random Forest classifier and are calculated per different experiment setup (i.e. per classification target, per test mode, per feature set).

To evaluate a classifier, we use the standard \textit{F1 score}~\cite{yan2017survey}, namely the harmonic mean of the Precision and Recall scores. These two metrics are normally a trade-off of each other, and a high F1 score indicates a good balance between these two metrics. Additionally, the training and testing times of the classifiers are assessed. 

The data is processed on a computing cluster with a Hadoop File System (HDFS) suitable for big data. The cluster consists of 55 nodes with each 32 GB RAM and 8-16 cores running Apache PySpark 2.2.0. The training and testing is done with the package Spark Sklearn \cite{spark_sklearn}, a distributed implementation of the machine-learning classifiers in the popular package machine-learning library Scikit-learn \cite{scikitlearn}. Spark Sklearn is adjusted to provide and log more information during the execution of experiments.

\section{Results}\label{sec:results}

\subsection{Test mode \textit{normal holdout}}

An overview of the performances of the classifiers for classification target \textit{all malware} and test mode \textit{normal holdout} is shown in Table \ref{tab:results_all}.
This table shows the feature categories present in the best model, i.e. with the best hyperparameter settings. From all models with similar performance as the best model of that classifier (performance results that do not differ statistically significantly (McNemar test a < 0.05 \cite{dietterich1998approximate}), the classifier with the least number of features is chosen, because this shows which features are relevant for detecting specific malware types. The number of features is based on the RFCV method described in Section \ref{sec:method}. Table \ref{tab:results_all} shows the following:

\begin{enumerate}
\itemsep-.2em
\item RF has the highest F1 score of 0.73 with feature set Apps and 0.72 with feature set Combined. 
\item All classifiers have an F1 score below 0.42 with feature set Global. This suggests that global features are less discriminative when detecting malicious actions.
\item RF with feature set Combined uses 10 app features (related to: App CPU, App Memory, App Process) to detect malicious actions of malware, suggesting that 10 app features are sufficient in detecting malicious actions of mobile malware. These 10 features are shown in Table \ref{tab:features_overview_rf_all}. All features and their corresponding feature category are shown in Tables \ref{tab:features_description_and_category_1} and \ref{tab:features_description_and_category_2}.
\item All classifiers have a higher FNR than FPR. This indicates that most models incorrectly classify a malicious action as benign more often than vice versa.
\end{enumerate}

\begin{table}[htb]
\caption{Comparison performance per classifier for classification target \textit{all malware}}
\label{tab:results_all}
\begin{minipage}{\columnwidth}
    {
\begin{tabularx}{\textwidth}{lCCCCCCCCC}
\toprule
\textbf{Classifier} & \multicolumn{3}{c}{\textbf{AdaBoost}} & \multicolumn{3}{c}{\textbf{Random Forest}} & \multicolumn{3}{c}{\textbf{KNN}} \tabularnewline
\textbf{Feature set} & Global & Apps & Comb. & Global & \textbf{Apps} & Comb. & Global & Apps & Comb. \tabularnewline

\multicolumn{10}{l}{\textbf{Best trained model}}\tabularnewline
\cmidrule(lr){2-4}\cmidrule(lr){5-7}\cmidrule(lr){8-10}
Nr. features & 26 & 25 & 43 & 24 & \textbf{29} & 10 & 26 & 13 & 9\tabularnewline
Accuracy & 0.922 & 0.943 & 0.945 & 0.922 & \textbf{0.958} & 0.959 & 0.897 & 0.944 & 0.946\tabularnewline
F1 score & 0.226 & 0.595 & 0.610 & 0.422 & \textbf{0.730} & 0.722 & 0.356 & 0.674 & 0.688\tabularnewline
FPR & 0.003 & 0.012 & 0.013 & 0.022 & \textbf{0.013} & 0.009 & 0.048 & 0.029 & 0.028\tabularnewline
FNR & 0.868 & 0.522 & 0.502 & 0.670 & \textbf{0.346} & 0.380 & 0.674 & 0.336 & 0.320\tabularnewline
\multicolumn{10}{l}{\textbf{Feature categories used}}\tabularnewline
\cmidrule(lr){2-4}\cmidrule(lr){5-7}\cmidrule(lr){8-10}
Battery & \checkmark &  & \checkmark & \checkmark &  &  & \checkmark &  & \tabularnewline
CPU & \checkmark &  & \checkmark & \checkmark &  &  & \checkmark &  & \tabularnewline
IO Interrupts &  &  &  &  &  &  &  &  & \tabularnewline
Memory & \checkmark &  & \checkmark & \checkmark &  &  & \checkmark &  & \tabularnewline
Metadata &  &  &  &  &  &  &  &  & \tabularnewline
Network & \checkmark &  & \checkmark & \checkmark &  &  & \checkmark &  & \tabularnewline
Storage &  &  &  &  &  &  &  &  & \tabularnewline
Wifi &  &  &  &  &  &  &  &  & \tabularnewline
App CPU &  & \checkmark & \checkmark &  & \cmark & \checkmark &  & \checkmark & \checkmark\tabularnewline
App Info &  &  &  &  &  &  &  &  & \tabularnewline
App Memory &  & \checkmark & \checkmark &  & \cmark & \checkmark &  & \checkmark & \checkmark\tabularnewline
App Network traffic &  & \checkmark & \checkmark &  & \cmark &  &  & \checkmark & \tabularnewline
App Process &  & \checkmark & \checkmark &  & \cmark & \checkmark &  & \checkmark & \checkmark \tabularnewline
\bottomrule
\end{tabularx}
}
\end{minipage}
\end{table}

Table \ref{tab:performance_permwtype_overview} shows an overview of the performance per classifier per malware type and feature set. Following the same reason as for Table \ref{tab:results_all}, the performances are shown for the least number of features and the best hyperparameter settings per classifier. Table \ref{tab:performance_permwtype_overview} shows the following:

\begin{enumerate}
\itemsep-0.2em
\item RF has the highest F1 score for 5 out of 10 malware types.
\item KNN has the highest F1 score for 5 out of 10 malware types.
\item AdaBoost has the highest F1 score for 4 out of 10 malware types \footnote{Some performance scores are the same, leading to a draw, hence to more than expected highest F1 scores.}.
\item AdaBoost has a high F1 score (>0.7) for 8 out of 10 malware types.
\item RF has a high F1 score for 7 out of 10 malware types.
\item KNN has a high F1 score for 7 out of 10 malware types.
\item All classifiers have a low F1 score for malware type 4 (Spyware SMS) and 6 (Adware). 
\end{enumerate}

\begin{table}[htb]
\caption{Comparison performance per classifier for classification target \textit{single malware} and test mode \textit{normal holdout} \\ (The numerical malware types in the first column are those listed in Table~\ref{tab:malware_versions}.)}
\label{tab:performance_permwtype_overview}
\begin{minipage}{\columnwidth}
\begin{center}
\begin{tabularx}{\textwidth}{CCCCCCCCCC}
\toprule
 & \multicolumn{3}{c}{\textbf{AdaBoost}} & \multicolumn{3}{c}{\textbf{Random Forest}} & \multicolumn{3}{c}{\textbf{KNN}} \tabularnewline
\textbf{Malware} & \textbf{F1} & \textbf{FPR} & \textbf{FNR} &\textbf{ F1 }& \textbf{FPR} & \textbf{FNR} & \textbf{F1}& \textbf{FPR} & \textbf{FNR} \tabularnewline
\cmidrule(lr){0-0}\cmidrule(lr){2-4}\cmidrule(lr){5-7}\cmidrule(lr){8-10}
1 & \textbf{0.704} & \textbf{0.025} & \textbf{0.321} & 0.571 & 0.025 & 0.500 & 0.536 & 0.047 & 0.464\tabularnewline
2 & 0.944 & 0.003 & 0.085 & \textbf{0.956} & \textbf{0.001} & \textbf{0.077} & 0.945 & 0.003 & 0.077\tabularnewline
3 & 0.764 & 0.000 & 0.382 & \textbf{0.793} & \textbf{0.003} & \textbf{0.324} & 0.727 & 0.003 & 0.412\tabularnewline
4 & 0.400 & 0.029 & 0.667 & \textbf{0.500} & \textbf{0.000} & \textbf{0.667} & 0.400 & 0.029 & 0.667\tabularnewline
5 & 0.927 & 0.005 & 0.089 & 0.927 & 0.005 & 0.089 & \textbf{0.946} & \textbf{0.005} & \textbf{0.054}\tabularnewline
6 & 0.194 & 0.011 & 0.880 & 0.271 & 0.039 & 0.782 & \textbf{0.352} & \textbf{0.057} & \textbf{0.662}\tabularnewline
7 & 0.813 & 0.015 & 0.216 & \textbf{0.857} & \textbf{0.003} & \textbf{0.227} & 0.793 & 0.012 & 0.268\tabularnewline
8 & \textbf{1.000} & \textbf{0.000} & \textbf{0.000} & 0.800 & 0.000 & 0.333 & 0.857 & 0.032 & 0.000\tabularnewline
9 & 0.913 & 0.000 & 0.160 & \textbf{0.936} & \textbf{0.000} & \textbf{0.120} & \textbf{0.936} & \textbf{0.000} & \textbf{0.120}\tabularnewline
11 & 0.909 & 0.009 & 0.118 & \textbf{0.938} & \textbf{0.000} & \textbf{0.118} & \textbf{0.938} & \textbf{0.000} & \textbf{0.118}\tabularnewline
\cmidrule(lr){0-0}\cmidrule(lr){2-4}\cmidrule(lr){5-7}\cmidrule(lr){8-10}
avg. & \textbf{0.757} & \textbf{0.010} & \textbf{0.292} & 0.755 & 0.008 & 0.324 & 0.743 & 0.019 & 0.284\tabularnewline
stdev & 0.263 & 0.010 & 0.283 & 0.232 & 0.013 & 0.251 & 0.233 & 0.021 & 0.252\tabularnewline
\bottomrule
\end{tabularx}
\end{center}
\end{minipage}
\end{table}

\noindent The feature categories included in the feature sets of the best models of classifiers per malware type are shown in Table \ref{tab:results_permwtype_permwtype}. Following the same reason as for Table \ref{tab:results_all}, the performances are shown for the least number of features and the best hyperparameter settings per classifier. This shows the following:

\begin{enumerate}
\itemsep-0.2em
\item 6 out of 10 malware versions are best detected using only app features.
\item Malware version 4 (Spyware SMS) is best detected using only global features. However, note that the F1 score for Spyware SMS is the lowest of all malware versions. 
\item Malware version 11 (DOS) is best detected using only one global feature regarding network traffic.
\item Malware version 1 and 9 (Spyware contacts theft and Privilege Escalation / Hostile downloader) is best detected using both the Apps and Global features.
\item None of the classifier includes I/O interrupts, Storage, or Wifi features for any of the malware versions.
\end{enumerate}

\begin{table}[htb]
\caption{Best performing classifiers per malware type\\ (The numerical malware types in the first column are those listed in Table~\ref{tab:malware_versions}.)}
\label{tab:results_permwtype_permwtype}
\begin{minipage}{\columnwidth}
\begin{center}
\begin{tabular}{lcccccccccc}
\toprule
\multicolumn{1}{r}{\textbf{Malware}} & \textbf{1} & \textbf{2} & \textbf{3} & \textbf{4} & \textbf{5} & \textbf{6} & \textbf{7} & \textbf{8} & \textbf{9} & \textbf{11}\tabularnewline
\multicolumn{11}{l}{\textbf{Best trained model}}\tabularnewline
\midrule
Classifier & AdaBoost & KNN & RF & RF & KNN & KNN & RF & AdaBoost & RF & KNN\tabularnewline
Feature set & Comb. & Apps & Apps & Global & Apps & Apps & Apps & Comb. & Comb. & Global\tabularnewline
Nr. features & 49 & 8 & 5 & 28 & 7 & 11 & 9 & 14 & 10 & 1\tabularnewline
F1 score & 0.70 & 0.94 & 0.73 & 0.50 & 0.95 & 0.35 & 0.86 & 1.00 & 0.94 & 0.94\tabularnewline
\multicolumn{11}{l}{\textbf{Feature categories used}}\tabularnewline
\midrule
Battery & \checkmark &  &  & \checkmark &  &  &  &  & \checkmark & \tabularnewline
CPU & \checkmark &  &  & \checkmark &  &  &  &  & \checkmark & \tabularnewline
I/O Interrupts &  &  &  &  &  &  &  &  &  & \tabularnewline
Memory & \checkmark &  &  & \checkmark &  &  &  &  &  & \tabularnewline
Network traffic & \checkmark &  &  & \checkmark &  &  &  &  & \checkmark & \checkmark\tabularnewline
Storage &  &  &  &  &  &  &  &  &  & \tabularnewline
Wifi &  &  &  &  &  &  &  &  &  & \tabularnewline
App CPU & \checkmark & \checkmark & \checkmark &  & \checkmark & \checkmark & \checkmark & \checkmark & \checkmark & \tabularnewline
App memory & \checkmark & \checkmark & \checkmark &  & \checkmark & \checkmark & \checkmark & \checkmark & \checkmark & \tabularnewline
App network traffic & \checkmark &  &  &  &  & \checkmark &  &  &  & \tabularnewline
App process & \checkmark & \checkmark & \checkmark &  &  &  &  & \checkmark &  & \tabularnewline
\bottomrule
\end{tabular}

\end{center}
\end{minipage}
\end{table}

\subsection{Test mode \textit{unknown device}}

An overview of the performances of the classifiers for classification target \textit{all malware} and test mode \textit{unknown device} is shown in Table \ref{tab:results_all_unknown_device}. This table shows high False Negative Rates (FNR) for all classifiers. Therefore, we do not include the feature categories as in the previous sections, as little insight can be drawn from these poorly performing models. Table \ref{tab:performance_permwtype_overview_unknown_device} shows an overview of the performance per classifier per malware type and feature set. Comparing this Table with Table \ref{tab:performance_permwtype_overview}, we see the following:
\begin{enumerate}
\itemsep-0.2em
\item All classifiers show lower performance for 7 out of 10 malware versions (versions 1-4 and 7-9) for test mode \textit{unknown device}, compared to test mode \textit{normal holdout}. 
\item AdaBoost and RF show better performance for malware version 5 (Phishing) for test mode \textit{unknown device}, compared to test mode \textit{normal holdout}. 
\item AdaBoost shows a perfect score (f1 score of 1) for detecting malware version 8 (Ransomware) for test mode \textit{unknown device}. For test mode \textit{normal holdout}, AdaBoost showed a perfect score as well.
\item AdaBoost and KNN shows a better performance for malware version 11 (DOS) for test mode \textit{unknown device}, compared to test mode \textit{normal holdout}. 

\end{enumerate}

\begin{table}[htb]
\caption{Comparison performance per classifier for classification target \textit{All malware} and test mode \textit{unknown device}}
\label{tab:results_all_unknown_device}
\begin{minipage}{\columnwidth}
    {
\begin{tabularx}{\textwidth}{lCCCCCCCCC}
\toprule
\textbf{Classifier} & \multicolumn{3}{c}{\textbf{AdaBoost}} & \multicolumn{3}{c}{\textbf{Random Forest}} & \multicolumn{3}{c}{\textbf{KNN}} \tabularnewline
Feature set & Global & \textbf{Apps} & Comb. & Global & Apps & Comb. & Global & Apps & Comb. \tabularnewline

\multicolumn{10}{l}{\textbf{Best model}}\tabularnewline
\cmidrule(lr){2-4}\cmidrule(lr){5-7}\cmidrule(lr){8-10}
Nr. features & 29    & \textbf{28}    & 44    & 28    & 6     & 8     & 30    & 4     & 11 \\
Accuracy & 0.923 & \textbf{0.944} & 0.926 & 0.888 & 0.918 & 0.925 & 0.824 & 0.917 & 0.910 \\
F1 score & 0.207 & \textbf{0.583} & 0.563 & 0.249 & 0.562 & 0.561 & 0.202 & 0.537 & 0.522 \\
FPR   & 0.005 & \textbf{0.013} & 0.042 & 0.052 & 0.057 & 0.043 & 0.126 & 0.053 & 0.061 \\
FNR   & 0.878 & \textbf{0.528} & 0.423 & 0.776 & 0.362 & 0.423 & 0.732 & 0.415 & 0.407 \\
\bottomrule
\end{tabularx}
}
\end{minipage}
\end{table}

\begin{table}[htb]
\caption{Comparison performance per classifier for classification target \textit{single malware} and test mode \textit{unknown device}\\ (The numerical malware types in the first column are those listed in Table~\ref{tab:malware_versions}.)}
\label{tab:performance_permwtype_overview_unknown_device}
\begin{minipage}{\columnwidth}
\begin{center}
\begin{tabularx}{\textwidth}{CCCCCCCCCC}
\toprule
 & \multicolumn{3}{c}{\textbf{AdaBoost}} & \multicolumn{3}{c}{\textbf{Random Forest}} & \multicolumn{3}{c}{\textbf{KNN}} \tabularnewline
\textbf{Malware} & \textbf{F1} & \textbf{FPR} & \textbf{FNR} &\textbf{ F1 }& \textbf{FPR} & \textbf{FNR} & \textbf{F1}& \textbf{FPR} & \textbf{FNR} \tabularnewline
\cmidrule(lr){0-0}\cmidrule(lr){2-4}\cmidrule(lr){5-7}\cmidrule(lr){8-10}
1     & \textbf{0.444} & \textbf{0.028} & \textbf{0.600} & 0.364 & 0.056 & 0.600 & 0.316 & 0.153 & 0.400 \\
2     & 0.817 & 0.015 & 0.230 & \textbf{0.861} & \textbf{0.014} & \textbf{0.164} & 0.847 & 0.027 & 0.115 \\
3     & 0.618 & 0.013 & 0.514 & \textbf{0.712} & \textbf{0.013} & \textbf{0.400} & 0.597 & 0.080 & 0.343 \\
4     & 0.000 & 0.028 & 1.000 & 0.000 & 0.000 & 1.000 & 0.000 & 0.000 & 1.000 \\
5     & 0.967 & 0.000 & 0.064 & \textbf{0.978} & \textbf{0.000} & \textbf{0.043} & 0.932 & 0.000 & 0.128 \\
6     & 0.200 & 0.056 & 0.817 & \textbf{0.245} & \textbf{0.037} & \textbf{0.802} & 0.207 & 0.134 & 0.706 \\
7     & 0.480 & 0.027 & 0.500 & \textbf{0.583} & \textbf{0.019} & \textbf{0.417} & 0.476 & 0.015 & 0.583 \\
8     & \textbf{1.000} & \textbf{0.000} & \textbf{0.000} & 0.000 & 0.000 & 1.000 & 0.667 & 0.032 & 0.000 \\
9     & \textbf{0.901} & \textbf{0.037} & \textbf{0.089} & 0.889 & 0.037 & 0.111 & 0.871 & 0.022 & 0.178 \\
11    & \textbf{0.977} & \textbf{0.000} & \textbf{0.045} & 0.930 & 0.010 & 0.091 & 0.952 & 0.000 & 0.091 \\
\midrule
avg.  & \textbf{0.640} & \textbf{0.020} & \textbf{0.386} & 0.556 & 0.018 & 0.463 & 0.586 & 0.046 & 0.354 \\
stdev & \textbf{0.352} & \textbf{0.018} & \textbf{0.353} & 0.380 & 0.019 & 0.372 & 0.330 & 0.057 & 0.322 \\
\bottomrule
\end{tabularx}
\end{center}
\end{minipage}
\end{table}

\subsection{Training and testing times}
Table \ref{tab:train_test_times} shows the training and testing times per classifier. This shows that training times of both AdaBoost and RF are significantly higher than for KNN (an algorithm which does no training). In contrast, testing times for KNN are significantly higher in comparison.

\begin{table}[htb]
\caption{Training and testing times per classifier for classification target \textit{all malware} and test mode \textit{normal holdout}}
\label{tab:train_test_times}
\begin{minipage}{\columnwidth}
\begin{center}
\begin{tabularx}{\textwidth}{lCCCCCCCCC}
\toprule
\textbf{Classifier} & \multicolumn{3}{c}{\textbf{AdaBoost}} & \multicolumn{3}{c}{\textbf{Random Forest}} & \multicolumn{3}{c}{KNN} \tabularnewline
\textbf{Feature set} & Global & Apps & Comb. & Global & Apps & Comb. & Global & Apps & Comb. \tabularnewline

\cmidrule(lr){2-4}\cmidrule(lr){5-7}\cmidrule(lr){8-10}
\textbf{Train (s)} & 17 & 18 & 31 & <1 & 30 & 10.5 & <1 & <1 & <1 
\tabularnewline
\textbf{Test (s)}  & <1 & <1 & <1 & <1 & <1 & <1 & 2 & 3 & 8 
\tabularnewline
\bottomrule
\end{tabularx}

\end{center}
\end{minipage}
\end{table}

\section{Discussion}\label{sec:discussion}
Below a discussion on the main findings of this research is presented. Most related papers use a different evaluation method making direct comparison difficult. Nevertheless, we tried ordering the findings from good to bad, compared to existing detection methods.

\paragraph{High performance RF, KNN, AdaBoost for classification target \textit{single malware}}
In this research, RF and KNN showed good results (F1 scores on average above 0.7) when separate models were trained per malware type. These findings are in line with earlier studies that found good performances for RF \cite{alam_random_2013,ham2013analysis,canfora2016acquiring} and KNN \cite{saracino2016madam} on the detection of mobile malware. None of the literature we found examined the AdaBoost classifier, hence the good performance of AdaBoost is a new finding.

\paragraph{AdaBoost is consistent with different test modes}
All classifiers, except AdaBoost, show worse performances when tested on new devices (test mode \textit{unknown device}), than if tested on the same devices (test mode \textit{normal holdout}). Most studies do not provide information on the test mode, thus test modes of other research cannot be compared to our study. Our results suggest that AdaBoost may be more suitable when its purpose is to first build a detection method with a subset of users and subsequently use this detection method to protect other users. This approach is scalable: new users can be protected with an already existing detection method. Otherwise, the model needs to be retrained constantly for every new user. AdaBoost with the Apps feature set and test mode \textit{unknown device} had an F1 score of 0.583, an FPR of 0.013, and an FNR of 0.528. The high FNR calls for improvements in detecting malicious actions of new devices, given the better scalability of this method. Some suggestions are given in Section \ref{future_works}.

\paragraph{Perfect score on Ransomware, but little data}
AdaBoost showed a perfect score for malware version 8 (Ransomware). The dataset used for training and testing contained only 170 records for malware version 8. Given the low number of testing instances, it is hard to estimate whether this performance is the same on larger test sets.

\paragraph{High performance RF for classification target \textit{all malware}}
The best RF model with classification target \textit{all malware} used the Apps feature set, containing 29 app features, and had an F1 score of 0.730, an FPR of 0.013, and an FNR of 0.346 (i.e., TPR of 0.654). Similar results (with no statistically significant difference to the best RF model), are achieved with the Combined feature set (F1 score of 0.722, an FPR of 0.009, and TPR of 0.620). A TPR of 0.62 is relatively low compared to other studies on the dynamic detecting of mobile malware, which showed TPRs between 0.61 and 1 (see Section \ref{sec:related_works} for a complete comparison). However, in contrast to \cite{amos_applying_2013,milosevic2016malaware,canfora2016acquiring}, with TPRs between 0.82 and 0.97, our study used data from real-life users instead of virtual environments for the training and testing of detection methods. The use of real-life data for the training of models may have led to lower performance due to the additional noise included in our dataset (see paragraph \textit{Overall high false-negative rate} below). 

Other studies using real devices for training and testing of models, based their performance on apps running isolated for 10 minutes \cite{amos_applying_2013}, devices in an idle state \cite{caviglione_seeing_2016} or unknown circumstances \cite{ham2013analysis,attar_gaussian_2014}. One study that used real devices under real-life circumstances for the assessment of their detection method is \cite{saracino2016madam}. In that study, the researchers created a multi-level framework (called MADAM) that showed high performance (TPR 0.97, FPR 0.005). MADAM requires root permissions as it used System Calls for the detection of malware. In contrast to MADAM, we do not require any root permissions. Furthermore, MADAM used both dynamic and static features, while we used only dynamic features. Lastly, MADAM is a detection method consisting of multiple architectural blocks that monitor different aspects of the device, in contrast to this present study that does not consist of a complex architecture.

\paragraph{False-negative rate higher than false-positive rate}
In this research, the FNR (undetected malicious actions) was overall higher than the FPR (benign actions labelled as malicious). In other studies that we examined during our literature research, the FPR is overall higher than the FNR. This may be due to the difference in the distribution of malicious and benign data points. In our research, the dataset contained 90\% benign data points and 10\% malicious data points. As most models are biased towards the majority class, the FNR is expected to be higher than the FPR in our case. In other studies the distribution of malicious and benign data points used are often 50/50 or a majority of malicious data points, resulting in equal FPRs and FNRs, or higher FPRs than FNRs. This stresses the importance of presenting both FPR and FNR values in research, which was not done in all examined studies.

\paragraph{Testing times long for KNN}
The testing times for KNN were more than 3 seconds for both the Apps and Combined feature set. This is based on our experimental setup which contained more computing power than the average smartphone. Given that in real-life, malicious or suspicious data should be tracked and noticed near real-time, long testing times may not be practical or feasible. Additionally long testing times stress the battery more as long computations are needed. Therefore, based on testing times, AdaBoost and RF may be preferable when models are running on smartphones. 

\paragraph{Low predictability on Spyware SMS and Adware}
Performance scores on the detection of malware versions 4 (Spyware SMS) and 6 (Adware) were low (F1 score below 0.5). The dataset used for training and testing contained 190 records for malware version 4. The low performance on the detection of Spyware SMS may, therefore, be due to the small number of training instances. The training and testing dataset contained 7940 records for malware version 6. Therefore, the low performance is most likely not a result of insufficient training instances, and more research is needed to find the cause for the low performance on detecting Adware.

\paragraph{Lower F1-scores compared to other state-of-the-art dynamic malware detectors}
Our best classifier (RF) shows an F1-score of 0.73. This is lower compared to state-of-the-art dynamic malware detectors such as Cai \cite{cai2018droidcat} and Saracino \cite{saracino2016madam}.
In \cite{cai2018droidcat}, the performance is high, but the results are taken from malware running in virtual environments for only 10 minutes with pseudo-random user input from Monkey application. This may lower the confidence that the models would perform the same when evaluated on a real device with a real user.  Additionally, Cai \cite{cai2018droidcat} uses non-hardware features such as Method Calls for their detection method. Our paper focuses on hardware features only. Our lower performance may indicate that using hardware features only result in lower overall performance. Saracino \cite{saracino2016madam} uses a combination of hardware features and non-hardware features (System Calls) for their detection methods. As mentioned in Section \ref{sec:related_works}, System Calls require root permissions which our detection method does require. However, still the detection method by Saracino shows a higher performance, indicating again that a combination of hardware and non-hardware features may increase the performance of dynamic malware detectors.

\paragraph{Overall high false-negative rate}
In this research, all detection methods showed relatively high FNR (F1 score above 0.3). This implies that a high number of malicious actions are undetected, and may have different causes. Many features in the feature set are influenced by many factors, as the features describe devices that are running multiple applications simultaneously. For instance, the priority, the CPU allocation, and the memory allocation of the malware app depend on other applications running parallel to it. Therefore, the features in the dataset do not solely reflect the (type of) action of the malware app, but also the state of the device at a given moment. This may result in excessive noise in the sensed data. Another possible cause may the fact that the influences of malicious and benign actions on the features sensed are similar, and indistinguishable to the classifier.

\section{Limitations}\label{sec:limitations}

All detection methods in this research are created using the same dataset. The limitations caused by using this particular dataset are listed below: 

\begin{enumerate}
\itemsep-0.2em
\item All devices in the dataset are Samsung Galaxy S5. Therefore, it is unknown how the models perform when used on other devices.
\item All malware types in the dataset were written for the purpose of this research. This limits the findings of the research due to two reasons. First: before the malware probe sent any data to a server, the data was scrambled to ensure the privacy of the volunteers. It is possible that this scrambling influenced the features analysed in the research. As a result, our detection models may be biased towards detecting scrambling actions, limiting its efficacy on wild malware. Second, it is unknown whether real malware executes the same way as custom malware. Although the behaviour of custom malware is based on wild malware, its implementation may differ. It is therefore unknown whether similar detection performance is to be expected on wild malware. 
\item The dataset contained a low number of data points for malware versions 4 and 6. This limits the conclusions drawn from their results.
\item The dataset did not contain any information on the creation times of the malware apps. Therefore, the sustainability metric, as mentioned in recent works \cite{cai2019assessing}\cite{onwuzurike2019mamadroid}, could not be included in the analysis.  The papers argue argue that detection methods should be evaluated by taking creation time of apps into considerations. They argue that detection methods should be trained on older apps (e.g. 2018) and tested on newer apps (e.g. 2019). This way detection methods are evaluated on their ability to detect unseen behaviour from newer apps. 
\end{enumerate}

The detection methods of this research share characteristics that may limit their performance or applicability. The limitations imposed by this are listed below:

\begin{enumerate}
\itemsep0.2em
\item All detection methods in this research are signature-based, i.e., identify malware based on a pattern of behaviour. In this research, the signature of a malicious action is a record of feature values from a some given feature set. This signature may be different for other (wild) malware types. Therefore, it is unknown whether similar performances are achieved on other (wild) malware types or other versions of the same malware type.
\item All detection methods in this research use RFCV as a feature-selection method. This method allows for feature reduction but is a greedy-search strategy, so may be sub-optimal. 
\item All detection methods using Apps features require data collection of all applications running. In our research, we discarded data from other applications, as we knew which application was malicious. However, in real-life circumstances, the malware application is unknown. Therefore, although Apps features show better performance than Global features, they are more resource-intensive to monitor than Global features. A suggestion for this issue is described in \ref{future_works}. 
\end{enumerate}

\section{Conclusion}\label{sec:conclusion}

This research showed i) what machine-learning classifier is most suitable for detecting Mobile Trojans, and ii) which aspects of a smartphone (features), related to hardware, are most important in detecting Mobile Trojans. The Random Forest classifier showed the highest performance in F1-score, compared with AdaBoost and KNN,  when one model is used that is trained on many types of Mobile Trojans. This classifier achieves an F1-score of 0.73 with an FPR of 0.009 and an FNR of 0.380. Random Forest, AdaBoost and K-Nearest Neighbour show high performances when separate models are trained on each type of Mobile Trojan with an average F1-score above 0.72, FPR below 0.02 and FNR below 0.33. Additionally, on 5 malware types (Spyware, Phishing, Ransomware, Privilege Escalation, DOS) we show high performances (F1-score above 0.963 and FNR below 0.120) making these models relevant for future detection methods that need to detect these malware types. 

Multiple dynamic hardware feature sets are examined in this research, making a distinction between global device features and features related to an application. When using one model for the detection of multiple types of Mobile Trojans, 10 app features related to the memory usage, process information, and CPU usage of the app, are sufficient for detecting malicious actions. Additionally, app features, in general, showed the best performance for detecting 7 out of 10 malware versions, compared to global device features. 

Recent related work performed their research in lab-like environments with models based on emulators, simulated user input, and/or short testing durations. To create models that can detect mobile malware threats in real life, we need models assessed on data of real devices rather than data of lab-like environments. This research tried to fill this gap by using data of real devices gathered over multiple years with malware based on real malware.

\section{Future work}\label{future_works}

Some suggestion for future research are listed below:

\begin{enumerate}
\itemsep-.2em
\item Focus on improving False Negative Rates as this research's detection methods show relatively high FNR. As Random Forest and AdaBoost showed relative high performance, other (boosted) ensemble classifiers can be examined such as GradientBoosting \cite{friedman2001greedy}.
\item Improve the performance of detection methods on new devices and device models if new data becomes available.
\item Improve dynamic detection methods by analysing real wild malware samples to improve the efficacy of the methods on the detection of real malware to overcome the limitation imposed by the use of self-written malware. 
\item Increase the sample size to improve the efficacy and applicability of the methods. This research analysed 10 different samples of Mobile Trojans.  
\item Examine whether time series analysis may improve the detection methods. This research's detection methods analyse dynamic features by individually assessing these values, without considering these value prior in time. Therefore, the absolute values of features are analysed. With time series analysis, relative values can be used which may improve the detection methods, as is suggested by the results of
 \cite{milosevic2016malaware}. 
\item Optimize the resource requirements of detection methods using App features. As described in Section \ref{sec:discussion}, monitoring App features of all applications installed on a device is resource-intensive. Research to decrease these resource requirements is needed. A suggestion may be to only monitor applications that require sensitive permissions, e.g. access to SD card or access to contacts, as was implemented in the framework of \cite{saracino2016madam}.
\item Improve proposed detection methods by taking sustainability as an additional metric for detectors. Recent works \cite{cai2019assessing}\cite{onwuzurike2019mamadroid} have argued that detection methods should be evaluated by taking creation time of apps into considerations. They argue that detection methods should be trained on older apps (e.g. 2018) and tested on newer apps (e.g. 2019). This way detection methods are evaluated on their ability to detect unseen behaviour from newer apps. The malware applications of our research did not contain information on the creation time, hence this could not be evaluated. 

\end{enumerate}

\section{Acknowledgements}
The authors would like to thank the Ben-Gurion University, for providing this research's dataset.

\bibliographystyle{unsrt}  
\bibliography{main_journal.bib}

\newpage
\section{Appendix}\label{appendix}

\begin{table}[htb]
    \caption{All features in different feature sets (for the meaning of each feature, see the documentation of the original dataset \cite{sherlock})}
    \label{tab:features_used_overview}
    \begin{minipage}{\columnwidth}
    \begin{center}
\small
\begin{tabular}{ccc}
\toprule
\textbf{Global} & \textbf{Apps} & \textbf{Combined} \\
\midrule
userid & userid & Global plus Apps \\
uuid  & uuid  &  \\
version & applicationname &  \\
traffic\_mobilerxbytes & cpu\_usage &  \\
traffic\_mobilerxpackets & packagename &  \\
traffic\_mobiletxbytes & packageuid &  \\
traffic\_mobiletxpackets & uidrxbytes &  \\
traffic\_totalrxbytes & uidrxpackets &  \\
traffic\_totalrxpackets & uidtxbytes &  \\
traffic\_totaltxbytes & uidtxpackets &  \\
traffic\_totaltxpackets & cguest\_time &  \\
traffic\_totalwifirxbytes & cmaj\_flt &  \\
traffic\_totalwifirxpackets & cstime &  \\
traffic\_totalwifitxbytes & cutime &  \\
traffic\_totalwifitxpackets & dalvikprivatedirty &  \\
traffic\_timestamp & dalvikpss &  \\
battery\_charge\_type & dalvikshareddirty &  \\
battery\_current\_avg & guest\_time &  \\
battery\_health & importance &  \\
battery\_icon\_small & importancereasoncode &  \\
battery\_invalid\_charger & importancereasonpid &  \\
battery\_level & lru   &  \\
battery\_online & nativeprivatedirty &  \\
battery\_plugged & nativepss &  \\
battery\_present & nativeshareddirty &  \\
battery\_scale & num\_threads &  \\
battery\_status & otherprivatedirty &  \\
battery\_technology & otherpss &  \\
battery\_temperature & othershareddirty &  \\
battery\_timestamp & pgid  &  \\
battery\_voltage & pid   &  \\
cpuhertz & ppid  &  \\
cpu\_0 & priority &  \\
cpu\_1 & rss   &  \\
cpu\_2 & rsslim &  \\
cpu\_3 & sid   &  \\
total\_cpu & start\_time &  \\
totalmemory\_freesize & state &  \\
totalmemory\_max\_size & stime &  \\
totalmemory\_total\_size & tcomm &  \\
totalmemory\_used\_size & utime &  \\
\bottomrule
\end{tabular}%

    \end{center}
    \end{minipage}
\end{table}

\begin{table}[htb]
    \caption{Features included in best performing Random Forest models (classification target: all malware), ordered by decreasing feature importance (for the meaning of each feature, see the documentation of the original dataset \cite{sherlock})}
    \label{tab:features_overview_rf_all}
    \begin{minipage}{\columnwidth}
    \begin{center}
\small
\begin{tabular}{ccc}
\toprule
\textbf{Global} & \textbf{Apps} & \textbf{Combined} \\
\midrule
traffic\_mobilerxbytes & cpu\_usage\_mor\_app & dalvikprivatedirty\_mor\_app \\
traffic\_mobiletxbytes & uidrxbytes\_mor\_app & dalvikpss\_mor\_app \\
traffic\_mobiletxpackets & uidrxpackets\_mor\_app & importance\_mor\_app \\
traffic\_totalrxbytes & uidtxbytes\_mor\_app & num\_threads\_mor\_app \\
traffic\_totalrxpackets & uidtxpackets\_mor\_app & otherprivatedirty\_mor\_app \\
traffic\_totaltxbytes & cmaj\_flt\_mor\_app & otherpss\_mor\_app \\
traffic\_totaltxpackets & cstime\_mor\_app & rss\_mor\_app \\
traffic\_totalwifirxbytes & dalvikprivatedirty\_mor\_app & stime\_mor\_app \\
traffic\_totalwifitxbytes & dalvikpss\_mor\_app & utime\_mor\_app \\
traffic\_totalwifitxpackets & dalvikshareddirty\_mor\_app & vsize\_mor\_app \\
battery\_current\_avg & importance\_mor\_app &  \\
battery\_icon\_small & importancereasonpid\_mor\_app &  \\
battery\_level & lru\_mor\_app &  \\
battery\_temperature & nativeprivatedirty\_mor\_app &  \\
battery\_voltage & nativepss\_mor\_app &  \\
cpu\_0 & nativeshareddirty\_mor\_app &  \\
cpu\_1 & num\_threads\_mor\_app &  \\
cpu\_2 & otherprivatedirty\_mor\_app &  \\
cpu\_3 & otherpss\_mor\_app &  \\
total\_cpu & othershareddirty\_mor\_app &  \\
totalmemory\_freesize & pgid\_mor\_app &  \\
totalmemory\_max\_size & pid\_mor\_app &  \\
totalmemory\_total\_size & ppid\_mor\_app &  \\
totalmemory\_used\_size & priority\_mor\_app &  \\
      & rss\_mor\_app &  \\
      & start\_time\_mor\_app &  \\
      & stime\_mor\_app &  \\
      & utime\_mor\_app &  \\
      & vsize\_mor\_app &  \\
\bottomrule
\end{tabular}%

    \end{center}
    \end{minipage}
\end{table}

\begin{table}[htb]
    \caption{Features included in best performing AdaBoost models (classification target = all malware) ordered per feature importance (for the meaning of each feature, see the documentation of the original dataset \cite{sherlock})}
    \label{tab:features_overview_ada_all}
    \begin{minipage}{\columnwidth}
    \begin{center}
\small
\begin{tabular}{ccc}
\toprule
\textbf{Global} & \textbf{Apps} & \textbf{Combined} \\
\midrule
totalmemory\_total\_size & rss\_mor\_app & rss\_mor\_app \\
totalmemory\_used\_size & utime\_mor\_app & utime\_mor\_app \\
battery\_voltage & otherprivatedirty\_mor\_app & dalvikpss\_mor\_app \\
battery\_temperature & dalvikprivatedirty\_mor\_app & otherprivatedirty\_mor\_app \\
traffic\_totaltxbytes & importance\_mor\_app & importance\_mor\_app \\
total\_cpu & otherpss\_mor\_app & otherpss\_mor\_app \\
totalmemory\_freesize & dalvikpss\_mor\_app & dalvikprivatedirty\_mor\_app \\
traffic\_totalrxbytes & num\_threads\_mor\_app & num\_threads\_mor\_app \\
battery\_level & vsize\_mor\_app & vsize\_mor\_app \\
cpu\_0 & stime\_mor\_app & stime\_mor\_app \\
traffic\_totaltxpackets & dalvikshareddirty\_mor\_app & traffic\_totalrxbytes \\
cpu\_2 & cpu\_usage\_mor\_app & dalvikshareddirty\_mor\_app \\
cpu\_3 & uidtxbytes\_mor\_app & totalmemory\_used\_size \\
traffic\_totalwifitxbytes & othershareddirty\_mor\_app & cpu\_usage\_mor\_app \\
battery\_current\_avg & start\_time\_mor\_app & othershareddirty\_mor\_app \\
traffic\_totalrxpackets & pid\_mor\_app & total\_cpu \\
cpu\_1 & uidrxbytes\_mor\_app & uidtxbytes\_mor\_app \\
traffic\_mobilerxbytes & pgid\_mor\_app & totalmemory\_freesize \\
traffic\_totalwifirxbytes & lru\_mor\_app & traffic\_totaltxbytes \\
traffic\_mobiletxbytes & uidtxpackets\_mor\_app & start\_time\_mor\_app \\
totalmemory\_max\_size & ppid\_mor\_app & battery\_voltage \\
traffic\_totalwifitxpackets & nativeprivatedirty\_mor\_app & totalmemory\_total\_size \\
battery\_icon\_small & priority\_mor\_app & battery\_temperature \\
traffic\_mobiletxpackets & uidrxpackets\_mor\_app & pid\_mor\_app \\
traffic\_totalwifirxpackets & cstime\_mor\_app & uidrxbytes\_mor\_app \\
traffic\_mobilerxpackets &       & ppid\_mor\_app \\
      &       & battery\_level \\
      &       & cpu\_0 \\
      &       & totalmemory\_max\_size \\
      &       & traffic\_totalwifitxbytes \\
      &       & traffic\_totaltxpackets \\
      &       & cpu\_2 \\
      &       & cpu\_3 \\
      &       & lru\_mor\_app \\
      &       & uidtxpackets\_mor\_app \\
      &       & nativeprivatedirty\_mor\_app \\
      &       & priority\_mor\_app \\
      &       & cpu\_1 \\
      &       & traffic\_totalrxpackets \\
      &       & pgid\_mor\_app \\
      &       & traffic\_mobiletxbytes \\
      &       & cstime\_mor\_app \\
      &       & traffic\_totalwifirxbytes \\
\bottomrule
\end{tabular}%

    \end{center}
    \end{minipage}
\end{table}

\begin{table}[htb]
    \caption{Features included in best performing KNN models (classification target = all malware) ordered per feature importance (for the meaning of each feature, see the documentation of the original dataset \cite{sherlock})}
    \label{tab:features_overview_knn_all}
    \begin{minipage}{\columnwidth}
    \begin{center}
\small
\begin{tabular}{ccc}
\toprule
\textbf{Global} & \textbf{Apps} & \textbf{Combined} \\
\midrule
totalmemory\_total\_size & rss\_mor\_app & rss\_mor\_app \\
totalmemory\_used\_size & utime\_mor\_app & utime\_mor\_app \\
battery\_voltage & otherprivatedirty\_mor\_app & dalvikpss\_mor\_app \\
battery\_temperature & dalvikprivatedirty\_mor\_app & otherprivatedirty\_mor\_app \\
traffic\_totaltxbytes & importance\_mor\_app & importance\_mor\_app \\
total\_cpu & otherpss\_mor\_app & otherpss\_mor\_app \\
totalmemory\_freesize & dalvikpss\_mor\_app & dalvikprivatedirty\_mor\_app \\
traffic\_totalrxbytes & num\_threads\_mor\_app & num\_threads\_mor\_app \\
battery\_level & vsize\_mor\_app & vsize\_mor\_app \\
cpu\_0 & stime\_mor\_app &  \\
traffic\_totaltxpackets & dalvikshareddirty\_mor\_app &  \\
cpu\_2 & cpu\_usage\_mor\_app &  \\
cpu\_3 & uidtxbytes\_mor\_app &  \\
traffic\_totalwifitxbytes &       &  \\
battery\_current\_avg &       &  \\
traffic\_totalrxpackets &       &  \\
cpu\_1 &       &  \\
traffic\_mobilerxbytes &       &  \\
traffic\_totalwifirxbytes &       &  \\
traffic\_mobiletxbytes &       &  \\
totalmemory\_max\_size &       &  \\
traffic\_totalwifitxpackets &       &  \\
battery\_icon\_small &       &  \\
traffic\_mobiletxpackets &       &  \\
traffic\_totalwifirxpackets &       &  \\
traffic\_mobilerxpackets &       &  \\
\bottomrule
\end{tabular}%

    \end{center}
    \end{minipage}
\end{table}

\begin{table}[t]
\caption{All features and their feature category (part 1 of 2; for the meaning of each feature, see the documentation of the original dataset \cite{sherlock})}
\parbox{.45\linewidth}{
    \label{tab:features_description_and_category_1}
    \begin{tabular}[t]{ll}
\toprule
\textbf{Feature} & \textbf{Category} \\
\midrule
cguest\_time & App\_CPU \\
cpu\_usage & App\_CPU \\
cstime & App\_CPU \\
cutime & App\_CPU \\
guest\_time & App\_CPU \\
Itrealvalue & App\_CPU \\
nice  & App\_CPU \\
num\_threads & App\_CPU \\
priority & App\_CPU \\
Processor & App\_CPU \\
rt\_priority & App\_CPU \\
stime & App\_CPU \\
utime & App\_CPU \\
packagename & App\_Info \\
packageuid & App\_Info \\
version\_code & App\_Info \\
version\_name & App\_Info \\
cmaj\_flt & App\_Memory \\
cminflt & App\_Memory \\
dalvikprivatedirty & App\_Memory \\
dalvikpss & App\_Memory \\
dalvikshareddirty & App\_Memory \\
lru   & App\_Memory \\
majflt & App\_Memory \\
minflt & App\_Memory \\
nativeprivatedirty & App\_Memory \\
nativepss & App\_Memory \\
nativeshareddirty & App\_Memory \\
otherprivatedirty & App\_Memory \\
otherpss & App\_Memory \\
othershareddirty & App\_Memory \\
rss   & App\_Memory \\
rsslim & App\_Memory \\
vsize & App\_Memory \\
uidrxbytes & App\_Network \\
uidrxpackets & App\_Network \\
uidtxbytes & App\_Network \\
uidtxpackets & App\_Network \\
endcode & App\_Process \\
exit\_signal & App\_Process \\
Flags & App\_Process \\
importance & App\_Process \\
importancereasoncode & App\_Process \\
importancereasonpid & App\_Process \\
pgid  & App\_Process \\
pid   & App\_Process \\\bottomrule
\end{tabular}%
}
\parbox{.45\linewidth}{
\begin{tabular}[t]{ll}
\toprule
\textbf{Feature}   & \textbf{Category} \\
\midrule
ppid  & App\_Process \\
sid   & App\_Process \\
start\_time & App\_Process \\
startcode & App\_Process \\
state & App\_Process \\
tcomm & App\_Process \\
tgpid & App\_Process \\
Wchan & App\_Process \\
battery\_charge\_type & Battery \\
battery\_current\_avg & Battery \\
battery\_health & Battery \\
battery\_icon\_small & Battery \\
battery\_invalid\_charger & Battery \\
battery\_level & Battery \\
battery\_online & Battery \\
battery\_plugged & Battery \\
battery\_present & Battery \\
battery\_scale & Battery \\
battery\_status & Battery \\
battery\_technology & Battery \\
battery\_temperature & Battery \\
battery\_timestamp & Battery \\
battery\_voltage & Battery \\
btime & CPU \\
cpu\_0 & CPU \\
cpu\_1 & CPU \\
cpu\_2 & CPU \\
cpu\_3 & CPU \\
cpuhertz & CPU \\
ctxt  & CPU \\
processes & CPU \\
procs\_blocked & CPU \\
procs\_running & CPU \\
tot\_idle & CPU \\
tot\_iowait & CPU \\
tot\_irq & CPU \\
tot\_nice & CPU \\
tot\_softirq & CPU \\
tot\_system & CPU \\
tot\_user & CPU \\
total\_cpu & CPU \\
companion\_cpu0 & IO Interrupts \\
companion\_sum\_cpu123 & IO Interrupts \\
cpu123\_intr\_prs & IO Interrupts \\
cypress\_touchkey\_cpu0 & IO Interrupts \\
cypress\_touchkey\_sum\_cpu123 & IO Interrupts \\

\bottomrule
\end{tabular}%
}
\end{table}

\begin{table}[t]
\caption{All features and their feature category (part 2 of 2, for the meaning of each feature, see the documentation of the original dataset \cite{sherlock})}
\parbox{.49\linewidth}{
    \label{tab:features_description_and_category_2}
    \begin{tabular}{ll}
\toprule
\textbf{Feature} & \textbf{Category} \\
\midrule
flip\_cover\_cpu0 & IO Interrupts \\
flip\_cover\_sum\_cpu123 & IO Interrupts \\
function\_call\_interrupts\_cpu0 & IO Interrupts \\
function\_call\_interrupts\_sum\_cpu123 & IO Interrupts \\
home\_key\_cpu0 & IO Interrupts \\
home\_key\_sum\_cpu123 & IO Interrupts \\
msmgpio\_cpu0 & IO Interrupts \\
msmgpio\_sum\_cpu123 & IO Interrupts \\
pn547\_cpu0 & IO Interrupts \\
pn547\_sum\_cpu123 & IO Interrupts \\
sec\_headset\_detect\_cpu0 & IO Interrupts \\
sec\_headset\_detect\_sum\_cpu123 & IO Interrupts \\
slimbus\_cpu0 & IO Interrupts \\
slimbus\_sum\_cpu123 & IO Interrupts \\
synaptics\_rmi4\_i2c\_cpu0 & IO Interrupts \\
synaptics\_rmi4\_i2c\_sum\_cpu123 & IO Interrupts \\
volume\_down\_cpu0 & IO Interrupts \\
volume\_down\_sum\_cpu123 & IO Interrupts \\
volume\_up\_cpu0 & IO Interrupts \\
volume\_up\_sum\_cpu123 & IO Interrupts \\
wcd9xxx\_cpu0 & IO Interrupts \\
wcd9xxx\_sum\_cpu123 & IO Interrupts \\
active & Memory \\
active\_anon & Memory \\
active\_file & Memory \\
anonpages & Memory \\
buffers & Memory \\
cached & Memory \\
commitlimit & Memory \\
committed\_as & Memory \\
dirty & Memory \\
highfree & Memory \\
hightotal & Memory \\
inactive & Memory \\
inactive\_anon & Memory \\
inactive\_file & Memory \\
kernelstack & Memory \\
lowfree & Memory \\
lowtotal & Memory \\
mapped & Memory \\
Memfree & Memory \\
memtotal & Memory \\
mlocked & Memory \\
pagetables & Memory \\
shmem & Memory \\
slab  & Memory \\
\bottomrule
\end{tabular}%
}
\parbox{.49\linewidth}{
\begin{tabular}{ll}
\toprule
\textbf{Feature} & \textbf{Category} \\
\midrule
sreclaimable & Memory \\
sunreclaim & Memory \\
swapcached & Memory \\
swapfree & Memory \\
swaptotal & Memory \\
totalmemory\_freesize & Memory \\
totalmemory\_max\_size & Memory \\
totalmemory\_total\_size & Memory \\
totalmemory\_used\_size & Memory \\
unevictable & Memory \\
vmallocchunk & Memory \\
vmalloctotal & Memory \\
vmallocused & Memory \\
writeback & Memory \\
applicationname & Metadata \\
sherlock\_version & Metadata \\
userid & Metadata \\
uuid  & Metadata \\
traffic\_mobilerxpackets & Network \\
traffic\_mobiletxbytes & Network \\
traffic\_mobiletxpackets & Network \\
traffic\_timestamp & Network \\
traffic\_totalrxbytes & Network \\
traffic\_totalrxpackets & Network \\
traffic\_totaltxbytes & Network \\
traffic\_totaltxpackets & Network \\
traffic\_totalwifirxbytes & Network \\
traffic\_totalwifirxpackets & Network \\
traffic\_totalwifitxbytes & Network \\
traffic\_totalwifitxpackets & Network \\
external\_availableblocks & Storage \\
external\_availablebytes & Storage \\
external\_blockcount & Storage \\
external\_blocksize & Storage \\
external\_freeblocks & Storage \\
external\_freebytes & Storage \\
external\_totalbytes & Storage \\
internal\_availableblocks & Storage \\
internal\_availablebytes & Storage \\
internal\_blockcount & Storage \\
internal\_blocksize & Storage \\
internal\_freeblocks & Storage \\
internal\_freebytes & Storage \\
internal\_totalbytes & Storage \\
connectedwifi\_level & Wifi \\
connectedwifi\_ssid & Wifi \\
\bottomrule
\end{tabular}%
}
\end{table}

\end{document}